\documentclass[prl,showpacs,byrevtex,floatfix,twocolumn,superscriptaddress]{revtex4}
\usepackage{amssymb,graphicx,afterpage}
\begin{document}
\title{\boldmath Multicritical end-point of the first-order ferromagnetic transition in colossal magnetoresistive
manganites\unboldmath}
%
%
%
\author{L. Demk\'o}
\affiliation{Department of Physics, Budapest University of
Technology and Economics and Condensed Matter Research Group of the Hungarian Academy of Sciences, 1111 Budapest, Hungary} \affiliation{Spin
Superstructure Project, ERATO, Japan Science and Technology Agency
(JST), Tsukuba 305-8562, Japan}
\author{I. K\'ezsm\'arki}
\affiliation{Department of Physics, Budapest University of
Technology and Economics and Condensed Matter Research Group of the Hungarian Academy of Sciences, 1111 Budapest, Hungary} \affiliation{Spin
Superstructure Project, ERATO, Japan Science and Technology Agency
(JST), Tsukuba 305-8562, Japan} \affiliation{Department of Applied
Physics, University of Tokyo, Tokyo 113-8656, Japan}
\author{G. Mih\'aly}
\affiliation{Department of Physics, Budapest University of
Technology and Economics and Condensed Matter Research Group of the Hungarian Academy of Sciences, 1111 Budapest, Hungary}%
\author{N. Takeshita}
\affiliation{Correlated Electron Research Center (CERC), National
Institute of Advanced Industrial Science and Technology (AIST),
Tsukuba 305-8562}
\author{Y. Tomioka}
\affiliation{Correlated Electron Research Center (CERC), National
Institute of Advanced Industrial Science and Technology (AIST),
Tsukuba 305-8562}
\author{Y. Tokura}
\affiliation{Spin Superstructure Project, ERATO, Japan Science and
Technology Agency (JST), Tsukuba 305-8562, Japan}
\affiliation{Department of Applied Physics, University of Tokyo,
Tokyo 113-8656, Japan} \affiliation{Correlated Electron Research
Center (CERC), National Institute of Advanced Industrial Science and
Technology (AIST), Tsukuba 305-8562}
\date{\today}
%
%
\pacs{\ }
\begin{abstract}
We have studied the bandwidth--temperature--magnetic field phase
diagram of RE$_{0.55}$Sr$_{0.45}$MnO$_3$ colossal magnetoresistance
manganites with ferromagnetic metallic (FM) ground state. The
bandwidth (or equivalently the double exchange interaction) was
controlled both via chemical substitution and hydrostatic pressure with a
focus on the vicinity of the critical pressure
$p^{\ast}$ where
the character of the zero-field FM transition changes from first to
second order. Below $p^{\ast}$ the first-order FM transition extends
up to a critical magnetic field, $H_{cr}$. It is suppressed by
pressure and approaches zero on the larger bandwidth side where
the surface of the first-order FM phase boundary is terminated by a
multicritical end-point $(p^{\ast}$$\approx$$32$\,kbar, $T^{\ast}$$\approx$$188$\,K,
$H^{\ast}$$=$$0)$. The change in the character of the transition and
the decrease of the CMR effect is attributed to the reduced CO/OO
fluctuations.
\end{abstract}
\maketitle

Perovskite-type manganites exhibit various fundamental phenomena of
current interest including colossal magnetoresistance (CMR), photo-
and current-induced insulator to metal transition, first-order
ferromagnetic transition and gigantic magneto-electric effect
\cite{Dagotto02}. Most of them are
collective effects arising from the strong interplay among the
electronic degrees of freedom in the spin, charge and orbital sector
which are further coupled to the underlying lattice. One of the most
dramatic phase transformations induced by external stimuli is the
magnetic field driven paramagnetic insulator (PI) to FM transition.
It is observed in a broad range above the
Curie temperature and the huge resistivity change
associated with the transition is termed as colossal
magnetoresistance. In the context of the CMR effect and the orbital
order-disorder transition in manganites, the possibility that
the first-order nature of phase transitions in three dimensional
systems can be preserved in the presence of disorder has recently
attracted much interest \cite{Pradhan2007}.

Although the detailed mechanism of the CMR effect is still under
debate, some of the basic ingredients have already been clarified.
Among them the phase competition between the two robust neighboring
states, i.e. the charge- and orbital-ordered insulator (CO/OO) and
the ferromagnetic metal, is thought to be
indispensable \cite{Dagotto02,Murakami03}. In the
bandwidth--temperature ($w-T$) phase diagram of CMR mangnites with
low quenched disorder, these two phases are separated by a
first-order phase boundary which can extend to as high temperature
as $T\approx200$\,K (see Fig.~1). It is terminated by a bicritical point where
the CO/OO and FM transition line ($T_{CO}(w)$ and $T_C(w)$,
respectively) meet each other. The temperature-induced transition
from the high-temperature PI phase to the long-range ordered states on the both sides
close to the bicritical point is of first order \cite{Adams04,Kim02,Tomioka04}. If quenched disorder
is introduced to the lattice by alloying rare earth (RE) and alkali
earth (AE) atoms on the A-sites with different ionic radius, the phase diagram is
strongly modified; the bicritical point is suppressed and a
spin-glass state appears between the two ordered phases. On the
other hand, $T_{CO}(w)$ and $T_C(w)$ lines remain of first order for
a large variation of the bandwidth \cite{Tomioka04}.
\begin{figure}[h!]
\includegraphics[width=2.25in]{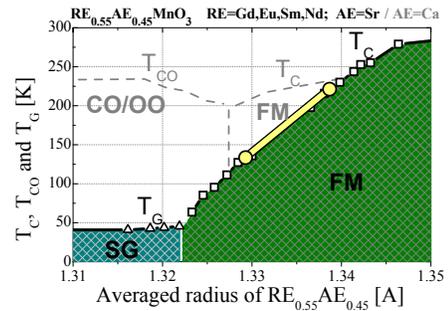}
\caption{(Color online) Bandwidth--temperature phase diagram of
RE$_{0.55}$AE$_{0.45}$MnO$_3$ manganites near half doping with high
level of quenched disorder (reproduced from Tomioka et al.
\cite{Tomioka04}). In contrast to the low-disorder case where
$T_{CO}$ and $T_C$ meet each other and form the so-called bicritical
point as represented by dashed lines, the ordered phases are suppressed and an intermediate spin
glass phase appears below $T_G$. The phase boundaries corresponding to
the high- and low-disorder case are shown for the two series, AE$=$Sr
and AE$=$Ca. In the present study the bandwidth
of the FM phase was controlled over the highlighted region by the
combination of chemical substitution and hydrostatic pressure.}
\label{fig1}
\end{figure}
\begin{figure}[ht!]
\includegraphics[width=3.45in]{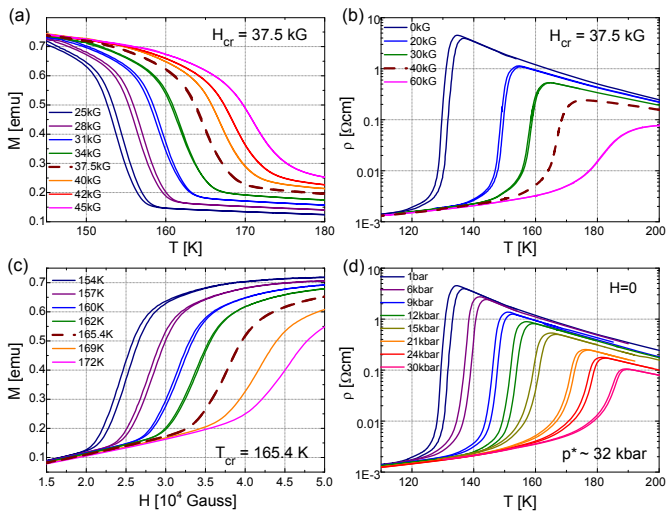}\\
\caption{(Color online) Mapping procedure of the
pressure-temperature-magnetic field phase diagram of
Sm$_{0.55}$Sr$_{0.45}$MnO$_3$ by magnetization and resistivity
measurements. Panel (a) and (b) show the temperature dependence of
the magnetization and resistivity, respectively, in various magnetic
fields. Above the critical field (temperature), $H_{cr}$ ($T_{cr}$) the hysteresis vanishes and the temperature-induced (field-induced) transition becomes a crossover. The corresponding curves are plotted by dashed lines. Panel (c) and (d) display the field dependence of the magnetization at various temperatures and the temperature dependence of
the resistivity at selected hydrostatic pressures, respectively.}
\end{figure}

This situation is shown in Fig.~1 where the typical $w-T$ phase
diagram of nearly half doped CMR manganites with large quenched
disorder is presented for the series of
RE$_{0.55}$Sr$_{0.45}$MnO$_3$ where RE$=$Gd, Eu, Sm, and Nd.
The variance of the ionic radius on the A-site, representing the level of disorder \cite{Attfield1998}, is $\sigma^2=\sum_ix_ir_i^2-r_A^2=0.01\pm0.003$\,{\AA}$^2$, where $x_i$, $r_i$, and $r_A$ are the fractional occupancies,
the ionic radii of the different cations, and the average ionic radius, respectively.
The first-order nature of the ferromagnetic phase
boundary is robust against the high level of disorder and also extends
over a large variation of the bandwidth. It stands from Eu$_{0.55}$Sr$_{0.45}$MnO$_3$
with the lowest transition temperature $T_C\approx50$\,K and changes the character somewhere between
Sm$_{0.55}$Sr$_{0.45}$MnO$_3$ and Nd$_{0.55}$Sr$_{0.45}$MnO$_3$ as it
becomes second order in the region of $T_C(w)=130-230$\,K.

In most ferromagnets, the transition from the
high-temperature disordered paramagnetic phase to the ferromagnetic
ground state is second order and characterized by a continuous
development of the magnetization below $T_C$. Furthermore, the
transition exists only in the zero-field limit and becomes a
crossover in presence of external magnetic field. On the other hand,
field-induced metamagnetic transitions between two ordered phases often
occur in a first-order manner accompanied with a discontinuous
change of the magnetization and a hysteresis.
Among a broad variety of such
systems \cite{Penc04}, the field-induced CO/OO
$\mapsto$ FM transition in CMR manganites is a representative
example. The first-order nature of the temperature- and field-induced PI $\mapsto$ FM transition
generally observed in CMR manganites -- either in the vicinity of
the bicritical point for low level of quenched disorder or next to
the spin glass phase when the bicritical
point is suppressed -- is likely to be the result of enhanced CO/OO fluctuations
which are present in a wide temperature range above the FM
state \cite{Murakami03,Adams04,Kim02,Tomioka04}.

In this study, we aim to map
and characterize the $T_C(w)$ phase boundary of (Sm$_{1-x}$Nd$_{x}$)$_{0.55}$Sr$_{0.45}$MnO$_3$
family combining the effect of chemical composition and hydrostatic pressure with a focus
on the highlighted region of Fig.~1 where the first-order transition changes to a second-order one. Both methods efficiently vary the effective one-electron bandwidth of these materials as either the increase of the average
ionic radius \cite{Tomioka03,Tomioka04,Arima95,Tomioka02,Brey05} or
the application of pressure \cite{Moritomo95,Neumeier95,Takeshita04}
causes the enhancement of the double exchange interaction responsible
for the FM phase. We also investigate the effect of
magnetic field on the FM transition and determine the $T_C(H)$ phase
boundary at each pressure. All the samples investigated here are single
crystals grown by a floating-zone method \cite{Tomioka03}. Resistivity measurements
were performed in the standard four-probe configuration and
magnetization was detected by using Quantum Design SQUID magnetometer. In both cases the application of hydrostatic pressure was carried
out using self-clamping pressure cells with kerosene as pressure-transmitting medium.

The dramatic increase of $T_C$ in presence of magnetic field is
characteristic of CMR manganites, especially in the presence of
quenched disorder. As shown in Fig.~2(a)-(b), the transition is
shifted to higher temperatures at an average rate of $\Delta
T_C(H)/\Delta H\approx0.9$\,K$/$kG$=9$\,K$/$T. Simultaneously, the
first-order nature of the transition weakens as clearly reflected by
the suppression of the magnetization and resistivity change
associated with the transition and also by the reduction of the hysteresis width. For the quantitative analysis we used the latter
quantity since it is more reliably obtained from the experimental data.
Though no detailed description about the effect of quenched disorder on the thermodynamic and transport properties
in the vicinity of a first-order transition is available,
we attribute the rounding-off in the resistivity and magnetization curves around $T_C$ to the influence of disorder on the nature of the transition \cite{note2}.
Above a critical magnetic field -- $H_{cr}\approx37.5$\,kG in case of Sm$_{0.55}$Sr$_{0.45}$MnO$_3$ --
the hysteresis completely vanishes and the first-order transition
becomes a crossover. (Throughout the paper $T_C$
stands for the zero-field transition temperature, unless field
dependence is explicitly indicated.) Similarly, by performing field
sweeps at fixed temperatures above $T_C$, the weakening of the
first-order character of the field-driven transition is discerned
with increasing temperature (see Fig.~2(c)). The hysteresis
vanishes above a critical temperature
$T_{cr}\approx165.4$\,K for Sm$_{0.55}$Sr$_{0.45}$MnO$_3$ at ambient
pressure. We call this ($H_{cr},T_{cr}$) point, the finite-field
critical end-point of the first-order FM transition.
\begin{figure}[t!]
\includegraphics[width=2.15in]{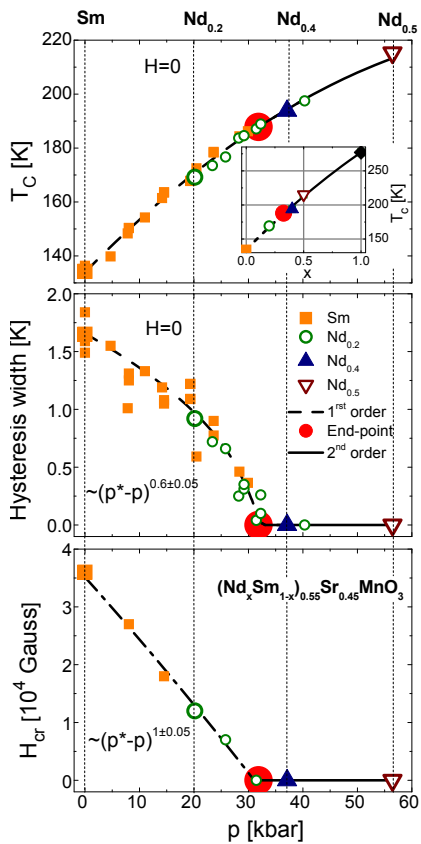}\\
\caption{(Color online) Top panel: Pressure dependence of the
zero-field ferromagnetic transition temperature in
(Sm$_{1-x}$Nd$_{x})_{0.55}$Sr$_{0.45}$MnO$_3$ ($x$$=$$0, 0.2, 0.4,$
and $0.5$). The inset displays $T_C$ as a function of chemical
composition $x$. The first- and second-order PI-FM phase boundary
are plotted by dashed and full line, respectively, while the
multicritical end-point separating them is labelled by a large full
circle. Middle panel: Zero-field
pressure dependence of the hysteresis width observed in the
temperature loops of the resistivity and magnetization curves.
Bottom panel: pressure dependence of the critical magnetic field
over the same series. The zero-pressure position of the different compounds is indicated
by vertical lines with  Sm$_{0.55}$Sr$_{0.45}$MnO$_3$ in the origin.}
\end{figure}
\begin{figure}[t!]
\includegraphics[width=2.25in]{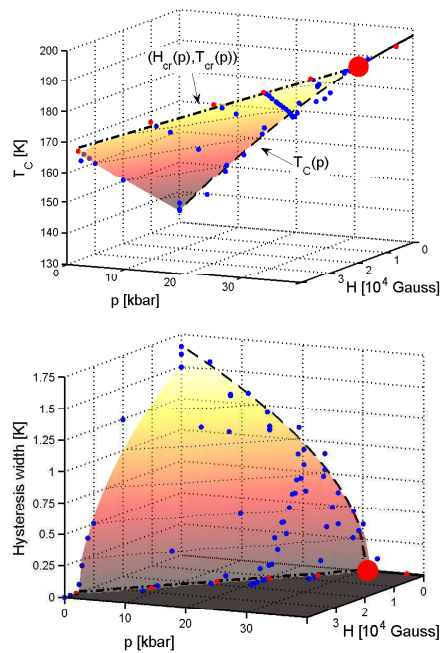}
\caption{(Color online) Upper panel: PM-FM transition over the
pressure--magnetic field--temperature phase diagram of
(Nd$_{x}$Sm$_{1-x})_{0.55}$Sr$_{0.45}$MnO$_3$ ($x=0, 0.2, 0.4,$ and
$0.5$). Dashed and solid lines represent the zero-field first-order and
second-order PM-FM transition, respectively. The dash-dotted line
corresponds to the finite-field critical end-line terminated by the
multicritical end-point labelled by large red circle.
The surface bordered by the dashed and dash-dotted lines
represent the first-order PI--FM phase boundary, as
obtained by interpolation of the data points indicated by blue dots.
Lower panel: Width of the temperature hysteresis over the
pressure--magnetic field plane for the same compounds. The origin of the
pressure ($p$) scales is taken at Sm$_{0.55}$Sr$_{0.45}$MnO$_3$.}
\end{figure}

The application of pressure increases the magnitude of the double
exchange interaction and consequently extends the FM phase to higher
temperatures -- $\Delta T_C(p)/\Delta
p\approx2$\,K$/$kbar as discerned in Fig.~2(d) -- while
the first-order nature of the transition is again reduced. The
critical pressure where the zero-field transition becomes of second
order is estimated to be $p^{\ast}\approx32$\,kbar for
Sm$_{0.55}$Sr$_{0.45}$MnO$_3$. In contrast to $T_{cr}(p)$ which is enhanced by
pressure in a parallel manner to $T_C(p)$, the critical magnetic field is suppressed according to
$H_{cr}(p)\propto (p^{\ast}-p)^{1\pm0.05}$ as indicated in Fig.~3. Above the critical pressure $p^{\ast}$ the finite-field FM transition
no longer exists. A systematic series of experiments, similar to those shown in Fig.~2,
were used to map the bandwidth--temperature--magnetic field phase
diagram for the full range of the compounds.

The evolution of the zero-field $T_C$, the hysteresis width
representing the first-order character of the transition, and the
critical field $H_{cr}$ as a function of pressure are shown in the respective panels of Fig.~3 in the
vicinity of the multicritical end-point located at
$(p^{\ast}$$\approx$$32$\,kbar, $T^{\ast}$$\approx$$188$\,K$)$. The
effect of pressure is nearly identical to that of the
chemical substitution as the pressure dependence of the
transition temperature for the different compounds fits to the same
$T_C(p)$ curve whose tendency is valid for the hysteresis width, as
well. We also determined the location of the critical point
expressed in terms of the relative composition of Nd and Sm and
found $x^{\ast}\approx0.33$. The whole
pressure--temperature--magnetic field phase diagram is displayed in
the upper panel of Fig.~4 while the lower panel shows the hysteresis
width over the corresponding area. The surface of the first-order
transition is bordered by two lines; the zero-field phase boundary
$T_C(p)$ and the finite-field critical end-line
($H_{cr}(p),T_{cr}(p)$). With increasing bandwidth the two lines
approach each other and the surface is terminated by the
multicritical end-point. On the larger bandwidth side of the
multicritical end-point a second-order FM transition line appears
where $T_C(p)$ still monotonically increases with pressure.

The first-order nature of the zero-field FM transition is the
strongest for Eu$_{0.55}$Sr$_{0.45}$MnO$_3$ with the lowest
transition temperature $T_C\approx50$\,K and largest critical field $H_{cr}\approx74$\,kG. The hysteresis width is
$\sim14.3$\,K and the resisitivity jump across the transition is
as large as nine orders of magnitude. This
compound is located on the verge of the SG-FM phase boundary and suffers the most from CO/OO fluctuations. Hence, it
shows the highest CMR effect among these compounds. The high-temperature
state cannot be regarded as a simple paramagnet with
disorder in the spin, charge and orbital sector as it is
essentially governed by short-range correlations of the CO/OO
phase \cite{Tomioka03,Motome03}. The finite correlation
length of the CO/OO fluctuations (typically a few lattice
constant just above $T_C$) can turn the generally continuous PI
$\mapsto$ FM transition to the first-order one.

Recent x-ray diffuse and Raman scattering measurements
\cite{Shimomura99,Shimomura00,Jirak00,Mathieu04,Tomioka03}
and optical conductivity experiments \cite{Kezsmarki08} have revealed
that the CO/OO correlations gradually vanish with increasing
temperature. Therefore, towards larger bandwidth, where $T_C$ is
remarkably enhanced, the phase fluctuations at the paramagnetic side
of the transition become less relevant. The CMR effect is suppressed
in accordance with the reduction of the resistivity change upon the
transition. Finally, at the multicritical end-point the
fluctuation-induced first-order transition is replaced by the usual
second-order FM transition. The colossal magnetoresistance does
still exist beyond this point but it is less effective due to the
rounding-off of the resistivity curves around $T_C$.

In conclusion, by combination of chemical and hydrostatic
pressure, we have studied the bandwidth--temperature--magnetic field
phase diagram of RE$_{0.55}$Sr$_{0.45}$MnO$_3$ colossal
magnetoresistance manganites with large quenched disorder. We have found that
the first-order nature of the PI $\mapsto$ FM transition is robust against
the A-site disorder. We have focused on the region where the character of the zero-field FM
transition changes from first to second order by variation of the bandwidth.
The hydrostatic pressure effectively control the
double exchange interaction as the transition is shifted to higher
temperatures at a rate of $\Delta T_C(p)/\Delta
p\approx2$\,K$/$kbar, while the first-order nature of the transition
weakens. Above a critical pressure
$p^{\ast}\approx32$\,kbar the zero-field FM transition becomes continuous.
In external magnetic field, the same tendencies are followed except for the
nature of the critical end-point. Although the first-order character
of the transition is similarly suppressed and eventually vanishes at
a critical magnetic field $H_{cr}(p)$, no phase transition exists
beyond this point. It becomes a crossover unlike in the zero-field
limit controlled by pressure. This finite-field critical end-line
approaches the $H$$=$$0$ plane according to $H_{cr}(p)\propto(p^{\ast}-p)^{1\pm0.05}$. Consequently, the
zero-field critical point $(p^{\ast},T^{\ast})$ is a multicritical
end-point terminating the surface of the first-order FM phase
boundary on the larger bandwidth side. The change in the character
of the transition and the weakening of the CMR effect are attributed
to the reduced CO/OO fluctuations.

The authors are grateful to N. Nagaosa and S. Onoda for enligthening
discussions. This work was supported by a Graint-In-Aid for
Scientific Research, MEXT of Japan and the Hungarian Scientific
Research Funds OTKA under grant Nos. F61413 and K62441 and Bolyai 00239/04.


\begin{references}
\bibitem{note2}Although the shape of the curves and the hysteresis width
were insensitive to the speed of either the temperature or magnetic field
sweeps ($v_T$ and $v_B$) we have performed all the experiments under uniform conditions,
namely with $v_T=0.2$\,K/min and $v_B=0.05$\,T/min.
\end{references}
\end{document}